\documentclass[twocolumn,showpacs,floatfix,preprintnumbers,amsmath,amssymb]{revtex4}

\usepackage{graphicx}
\usepackage{dcolumn}
\usepackage{bm}
\newcommand{\be}{\begin{equation}}
\newcommand{\ee}{\end{equation}}
\newcommand{\bea}{\begin{eqnarray}}
\newcommand{\beas}{\begin{eqnarray*}}
\newcommand{\eea}{\end{eqnarray}}
\newcommand{\eeas}{\end{eqnarray*}} 
\newcommand{\ba}{\begin{array}}
\newcommand{\ea}{\end{array}}

\begin{document}

\title{Finding an Upper Limit in the Presence of Unknown Background}
\affiliation{Department of Physics, 
                  University of California, Santa Barbara,   
	          Santa Barbara, CA 93106, USA   } 
\author{S.~Yellin}
\email[email:]{yellin@slac.stanford.edu}
\affiliation{Department of Physics, 
                  University of California, Santa Barbara,   
	          Santa Barbara, CA 93106, USA   }
\date{\today}
\begin{abstract}
Experimenters report an upper limit if the signal they are trying to detect
is non-existent or below their experiment's sensitivity.  Such experiments
may be contaminated with a background too poorly understood to subtract.
If the background is distributed differently in some parameter from the
expected signal, it is possible to take advantage of this difference to get
a stronger limit than would be possible if the difference in distribution
were ignored.  We discuss the ``maximum gap'' method, which finds the best
gap between events for setting an upper limit, and generalize to the ``optimum
interval'' method, which uses intervals with especially few events.  These
methods, which apply to the case of relatively small backgrounds, do not use
binning, are relatively insensitive to cuts on the range of the parameter,
are parameter independent (i.e., do not change when a one-one change of
variables is made), and provide true, though possibly conservative,
classical one-sided confidence intervals.
\end{abstract}
\maketitle

\section{Introduction}

Suppose we have an experiment whose events are distributed
along a one-dimensional interval.  The
events are produced by a process for which the expected shape of the
event distribution is known, but with an unknown normalization.
In addition to the signal, there may also be a background whose
expectation value per unit interval is known, but one cannot completely
exclude the possibility of an additional
background whose expectation value per unit interval is non-negative, but is
otherwise unknown.  If the experimenters cannot exclude the possibility that
the unknown background is large enough to account for all the events, they can
only report an upper limit on the signal.
Even experimenters who think they understand a background well enough to
subtract it may wish to allow for the possibility that they are mistaken by
also presenting results without subtraction.
Methods based on likelihood, such as the approach of
Feldman-Cousins~\cite{FeldCous}, or Bayesian analysis, cannot be applied
because the likelihood associated with an unknown background is unknown.
An example of this situation is analysis of an experiment which tries to detect
recoil energies, $E_{\mathrm{recoil}}$, deposited by WIMPs bouncing off atoms
in a detector.  For a given WIMP mass, and assumed WIMP velocity distribution,
the shape of the distribution in $E_{\mathrm{recoil}}$ can be computed, but the
WIMP cross section is unknown, and it is hard to be certain that all
backgrounds are understood.  The simplest way of dealing with such a situation
is to pick an interval in, say, $E_{\mathrm{recoil}}$, and take as upper limit
the largest cross section that would have a significant probability, say 10\%,
of giving as few events as were observed, assuming all observed events were
from WIMPs.  One problem with this naive method is that it can be very
sensitive to the interval chosen.  It is typical for the bottom of a
detector's range of sensitivity to be limited by noise or other backgrounds.
Thus if the interval extends to especially low $E_{\mathrm{recoil}}$,
there will be
many events, leading to a weaker (higher) upper limit than is required
by the data.  On the other hand, experimenters could inadvertently bias
the result by choosing the interval's endpoints to give especially
few events, with an upper limit that is lower than is justified by the data.
In order to avoid such a bias, it might be thought best to avoid using the
observed events to select the interval used.  But the procedures discussed here
take the opposite approach.
The range is carefully chosen to include especially few events compared with
the number expected from a signal.  The way the range is
chosen makes the procedure especially insensitive to
unknown background, which tends to be most harmful where there are
especially many events compared with the number expected from a signal.
It would be a mistake to compute the upper limit as if the interval were
selected without using the data; so the computation is designed to be correct
for the way the data are used.

While the
methods described here cannot be used to identify a positive detection, they
are appropriate for obtaining upper limits from experiments whose backgrounds
are very low, but non-zero.  These methods have been used by the CDMS
experiment~\cite{CDMS}.

\section{Maximum Gap Method}

\setkeys{Gin}{width=6 in} 
\begin{figure*}  
\includegraphics{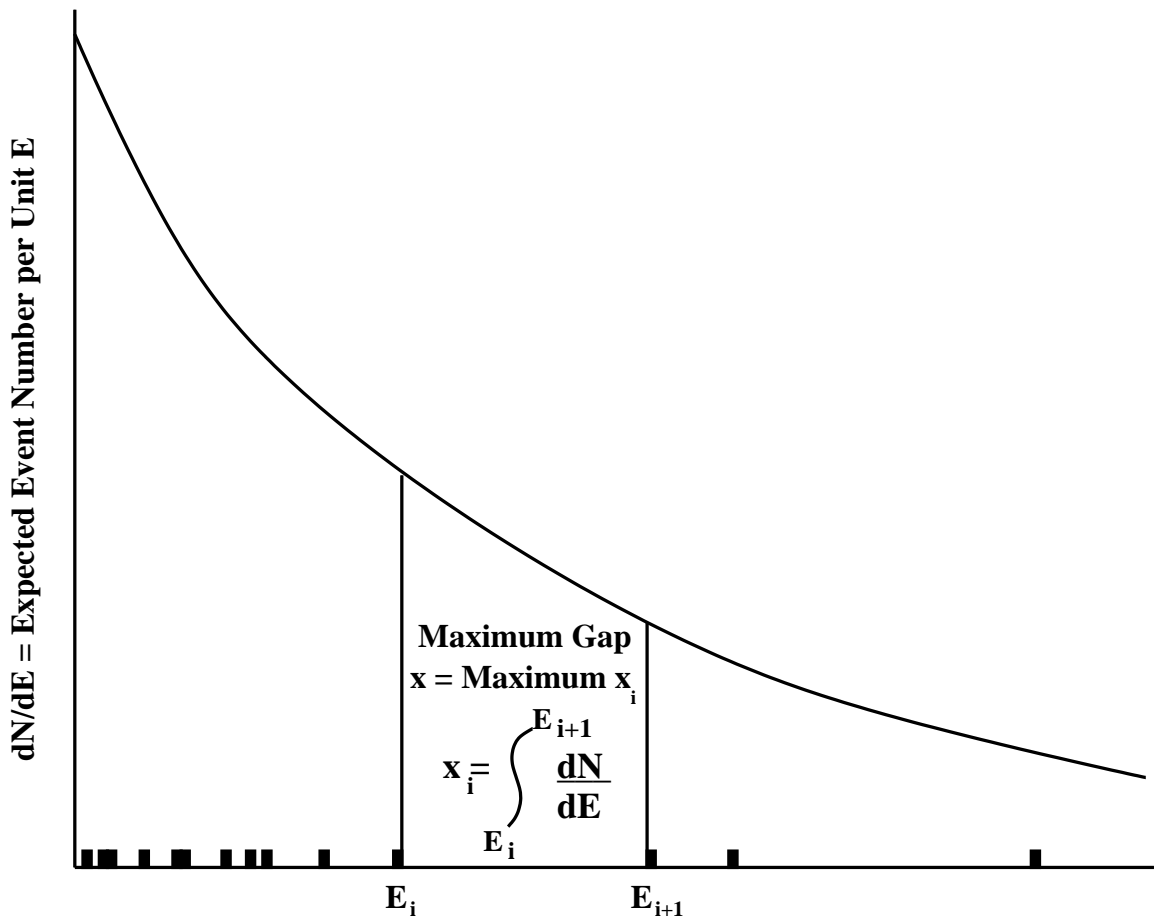}
\caption{Illustration of the maximum gap method.  The horizontal axis is
some parameter, ``$E$'', measured for each event.  The smooth curve is the signal
expected for the proposed cross section,
including any known background.  The events from signal,
known background, and unknown background are the small rectangles along the
horizontal axis.   The integral of the signal between two events is
``$x_{\mathrm i}$''.}
\label{maxgapfig}
\end{figure*}

\setkeys{Gin}{width=3.25 in} 
Figure \ref{maxgapfig} illustrates the maximum gap method.  Small
rectangles along the horizontal axis represent events, with position on
the horizontal axis representing some measured parameter, say ``energy'', $E$.
The curve shows the event spectrum, $dN/dE$, expected from a proposed
cross section, $\sigma$. If there is
a completely known background, it is included in $dN/dE$.  But whether
or not there is a completely known background, we assume there is also an
unknown background contaminating the data.
To set an upper limit, we vary the proposed size of $\sigma$
until it is just high enough to be rejected as being too high.  We seek a
criterion for deciding if a proposed signal is too high.
Since there are especially many events at low E, while $dN/dE$ is not
especially high there, those events must be mostly from
the unknown background.  If we only looked at the low energy part of the
data, we would have to set an especially weak (high) upper limit.  To find the
strongest (lowest) possible upper limit, we should look at energies where
there aren't many events, and therefore isn't much background.

Between any two events, $E_{\mathrm{i}}$ and $E_{\mathrm{i+1}}$, there is a
gap.  For a given value of $\sigma$, the
``size'' of the gap can be characterized by the value within the gap of
the expected number of events,
\be
x_{\mathrm{i}} = \int_{E_{\mathrm{i}}}^{E_{\mathrm{i+1}}} \frac{dN}{dE} dE.
\ee
The ``maximum gap'' is the one with the greatest ``size''; it is the largest
of all the $x_{\mathrm{i}}$. The bigger we assume
$\sigma$ to be, the bigger will be the size of the maximum gap in the
observed event distribution.  If we want, we can choose $\sigma$ so large
that there are millions of events expected in the maximum gap.  But such a
large $\sigma$ would be experimentally excluded, for unless a mistake has been
made, it is almost impossible to find zero events where millions are expected.
To express this idea in a less extreme form, a particular choice of $\sigma$
should be rejected as too large if, with that choice of
$\sigma$, there is a gap between adjacent events with ``too many'' expected
events.  The criterion for ``too many'' is that if the choice of $\sigma$
were correct, a random experiment would almost always give fewer expected
events in its maximum gap.  Call $x$ the size of the maximum gap in the
random experiment.
If the random $x$ is lower than the observed maximum gap size with
probability $C_0$, the
assumed value of $\sigma$ is rejected as too high with confidence level $C_0$.
Since $x$ is unchanged under a one-one transformation of the variable in
which events are distributed, one may
make a transformation at a point from whatever variable is used,
say $E$, to a variable equal
to the total number of events expected in the interval between the point and
the lowest allowed value of $E$.
No matter how events were expected to be distributed in the original variable,
in the new variable they are distributed uniformly with unit density.
Thus any event distribution is equivalent to a uniform distribution of unit
density.  The probability distribution of $x$ depends on the total length of
this uniform unit density distribution, and in this new variable the total
length of the distribution is equal to the total expected number of events,
$\mu$, but it does not depend on the shape of the original event distribution.
$C_0$, the probability of the maximum gap size being smaller than a particular
value of $x$, is a function only of $x$ and $\mu$:
\be
C_0(x,\mu) = \sum_{k=0}^m \frac {(kx-\mu)^ke^{-kx}}{k!}
\left(1 + \frac{k}{\mu-kx}\right), \label{C0eqn}
\ee
where $m$ is the greatest integer $\le\mu/x$.
For a 90\% confidence level upper limit, increase $\sigma$ until $\mu$
and the observed $x$ are such that $C_0$ reaches 0.90.

Equation \ref{C0eqn} can be evaluated relatively quickly when
$C_0$ is near 0.9.  When $\mu$ is small, so is $m$, and when
$\mu$ is large, the series can be truncated at relatively small $k$ without
making a significant error.  Equation \ref{C0eqn} is derived in Appendix A.

While this method can be used with an arbitrary number of events in the
data, it is most appropriate when there are only a few events in
the part of the range that seems relatively free of background (small $\mu$).
The method is not dependent on a
choice for binning because unbinned data are used.  No Monte Carlo
computation of the confidence level is needed because the same formula for
$C_0$ applies independent of the functional form for the shape of the expected
event distribution.  The result is a conservative upper limit that is not too
badly weakened by a large unknown
background in part of the region under consideration; the method
effectively excludes regions where a large unknown background
causes events to be too close together for the maximum gap to be there.

\section{Optimum Interval Method}

If there is a relatively high density of events in the data, we may want to 
replace the ``maximum gap'' method by one in which we consider,
for example, the ``maximum'' interval over which there is 1 event observed,
or 2 events, or $n$ events, instead of the zero events in a gap.

Define $C_n(x,\mu)$ to be the probability, for a given
cross section without background, that all intervals with $\le n$
events have their expected number of events $\le x$.  As for $C_0$ of
the maximum gap method, so long as $x$ and $\mu$ are fixed, $C_n$
is independent of the shape of the cross section and the parameter in which
events are distributed.  But $C_n(x,\mu)$ increases when $x$ increases, and
it increases when $n$ decreases.  $C_n$ can be tabulated with the
help of a Monte Carlo program, although the special case of $n=0$
can be more accurately computed with Eq.~\ref{C0eqn}.
Once $n$ is chosen, $C_n$ can be used in the same way as $C_0$ for obtaining
an upper limit: for $x$ equal to the maximum expected number of events taken
over all intervals with $\le n$ events,
$C_n(x,\mu)$ is the confidence level with which the assumed cross section
is excluded as being too high.  But since we do not want to allow $n$ to be
chosen in a way that skews results to conform with our prejudices, the optimum
gap method includes automatic selection of which $n$ to use.

For each interval within the total range of an actual experiment, compute
$C_n(x,\mu)$ for the observed number of events, $n$, and expected number
of events, $x$, in the interval.
The bigger $C_n$ is, the stronger will be the evidence that the assumed
cross section is too high.  Thus for each
possible interval, one may quantify how strongly the proposed cross section
is excluded by the data.  The ``optimum interval'' is the
interval that most strongly indicates that the proposed cross section is too
high.  The optimum interval tends to be one
in which the unknown background is especially small.  The overall test
quantity used for finding an upper limit on the cross section is then
$C_{\mathrm{Max}}$, the maximum over all possible intervals of $C_n(x,\mu)$.
A 90\% confidence level upper limit on the cross section is one 
for which the observed $C_{\mathrm{Max}}$ is higher than would be expected from
90\% of random experiments with that cross section and no unknown background.

The definition of $C_{\mathrm{Max}}$ seems to imply that its determination
requires checking an infinite number of intervals.
But given any interval with $n$ events, $x$, hence $C_n(x,\mu)$, can be
increased without increasing $n$ by expanding the interval until it almost
hits either another event or an endpoint of the total experimental range.
For determination of $C_{\mathrm{Max}}$ one need only consider intervals that are
terminated by an event or by an endpoint of the total experimental range.
If the experiment has $N$ events, then there are $(N+1)(N+2)/2$ such
intervals, one of which has $C_n(x,\mu)=C_{\mathrm{Max}}$.

\begin{figure}
\includegraphics{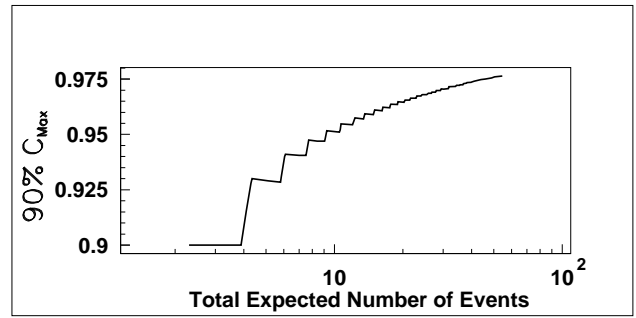}
\caption{Plot of $\bar C_{\mathrm{Max}}(.9,\mu)$, the value of
$C_{\mathrm{Max}}$ for which the 90\% confidence level is reached, as a
function of the total number of events $\mu$ expected in the experimental
range.}
\label{CMaxfig}
\end{figure}

The function $\bar C_{\mathrm{Max}}(C,\mu)$ is defined to be the value such that
fraction $C$ of random experiments with that $\mu$, and no unknown
background, will give $C_{\mathrm{Max}}<\bar C_{\mathrm{Max}}(C,\mu)$.  Thus the 90\%
confidence level upper limit on the cross section is where $C_{\mathrm{Max}}$ of the
experiment equals $\bar C_{\mathrm{Max}}(.9,\mu)$, which is plotted in
Fig.~\ref{CMaxfig}.

A Monte Carlo program was used to tabulate $C_n(x,\mu)$.  A
Fortran routine interpolates the table to compute
$C_n(x,\mu)$ when $n$, $x$, and $\mu$ are within the tabulated range.  The
routine applies when $0<\mu<54.5$ and when $0\le n\le 50$.

The function, $\bar C_{\mathrm{Max}}(C,\mu)$, has been
computed by Monte Carlo and tabulated for $\mu<54.5$ and various $C$.
Certain peculiarities of this function are discussed in Appendix B.

Routines to
evaluate functions described in this paper, along with tables they use, are
available on the web~\cite{software}.

\section{Comparisons of the Methods}

\begin{figure}
\includegraphics{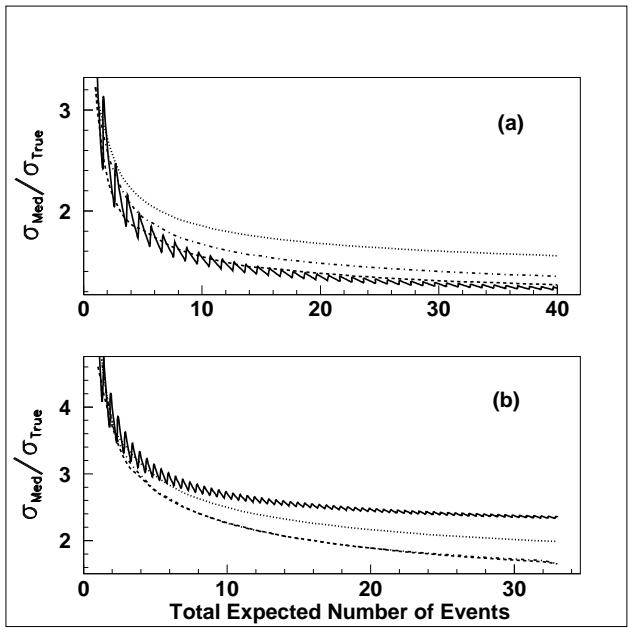}
\caption{$\sigma_{\mathrm{Med}}/\sigma_{\mathrm{True}}$, the typical factor by which the upper
limit cross section exceeds the true cross section, when $C_0$ is used
(dotted lines), when $p_{\mathrm{Max}}$ is used
(dash-dotted lines), when $C_{\mathrm{Max}}$ is used (dashed lines), and when the
Poisson method is used (solid lines).  These ratios are
a function of $\mu$, the total number of events expected from the true
cross section in the entire experimental range.  For the upper figure (a)
there is no background, and for the lower figure (b) there is just as much
unknown background as there is signal, but the background is concentrated in
a part of the experimental range that contains only half the total signal.
}
\label{comparefig}
\end{figure}

We compare the maximum gap ($C_0$) and optimum interval ($C_{\mathrm{Max}}$)
methods with each other, with the standard~\cite{PDG} way of finding an upper limit
(``Poisson''), and with another method ($p_{\mathrm{Max}}$) described in
Appendix C.

The standard ``Poisson'' confidence level $C$ upper limit
cross section is the one whose $\mu$ would result in fraction $C$ of random
experiments having more events in the entire experimental range
than the $n$ actually observed.  This fraction $C$ is
\be P(\mu,n+1)
\equiv \sum_{k=n+1}^\infty \frac
{\mu^k}{k!}e^{-\mu}
 = \int_0^\mu dt \frac{t^n}{n!} e^{-t}.
\label{PoisProb} \ee
The last equality is proved by observing that both sides have the same
derivative, and they have the same value at $\mu=0$.  $P(x,a)$, the incomplete
Gamma function, is in CERNLIB~\cite{CERNLIB} as GAPNC(a,x), DGAPNC(a,x),
and GAMDIS(x,a).

The description of the $p_{\mathrm{Max}}$ method is relegated to Appendix C
because although $p_{\mathrm{Max}}$ is somewhat easier to implement than
$C_{\mathrm{Max}}$, it was found to be less powerful.

Two comparisons of the effectiveness of the methods were performed: tests
``(a)'' and ``(b)''.
For test (a), $500,000$ zero-background Monte Carlo experiments were generated
for each of 40 assumed cross sections.  $C_0$, $p_{\mathrm{Max}}$, $C_{\mathrm{Max}}$, and the
Poisson method were used to find the 90\% confidence level upper limits on
the cross section.  For a given true cross section, $\sigma_{\mathrm{True}}$, there is
a certain median value, $\sigma_{\mathrm{Med}}$, that is exceeded exactly 50\% of the
time by the computed upper limit.  Fig.~\ref{comparefig}(a) shows
$\sigma_{\mathrm{Med}}/\sigma_{\mathrm{True}}$ as a function of $\mu$.  The dotted curve
used $C_0$ to determine the upper limit, the dash-dotted curve used
$p_{\mathrm{Max}}$, the dashed one used $C_{\mathrm{Max}}$, and the solid, jagged, curve
used the Poisson method.  The Poisson method gives a jagged curve because
of the discrete nature of the variable used to calculate the upper limit,
the total number of detected events.  For any cross section shape, when
there is no background, $C_{\mathrm{Max}}$ gives a stronger limit than
$p_{\mathrm{Max}}$ in most random experiments, and both are stronger than
$C_0$.  Even without background, for some values of the true $\mu$,
$C_{\mathrm{Max}}$ gives a stronger (lower) upper limit than the Poisson method.
This happens because the discrete nature of the Poisson method causes it to
have greater than 90\% coverage.

Although test (a) is presented as a comparison
of methods in the absence of background, it can also be considered to be a
comparison of methods when the background is distributed the same as the
signal.  If the unknown background happens to have the same distribution
as the signal would have, essentially no sensitivity is lost by using the
optimum interval method with $C_{\mathrm{Max}}$ instead of the Poisson method. 

Test (b) was similar to test (a), but the Monte Carlo program simulated
a background unknown to the experimenters, and distributed differently from
the expected signal.  The
total experimental region was split into a high part and a low part, with
background only in the low part.  Half the expected signal was placed in
the low part, where the simulated background was twice the
expected signal.  For this case, the two lowest curves are almost exactly
on top of each other; Fig.~\ref{comparefig}(b) shows that $C_{\mathrm{Max}}$
and $p_{\mathrm{Max}}$ get equally strong upper limits.  $C_0$ produces
a weaker limit, and the Poisson method is weakest of all.

\begin{figure}
\includegraphics{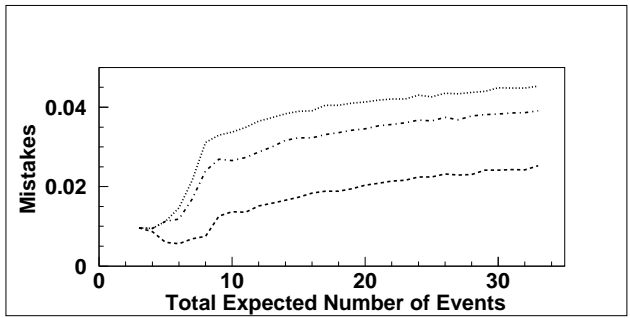}
\caption{Fraction of cases for test (b) (see text) in which the true
cross section was higher than the upper limit on the cross section
computed using $C_0$ (dotted), $p_{\mathrm{Max}}$ (dash-dotted) and
$C_{\mathrm{Max}}$ (dashed).
}
\label{mistakefig}
\end{figure}

From the definition of the 90\% confidence level upper limit, test (a)
results in an upper limit that is lower than the true value exactly 10\% of the
time; i.e., all methods except the Poisson make a mistake 10\% of the time
(the discrete nature of the Poisson distribution results in its making
mistakes less than 10\% of the time).
But for test (b), the unknown background raises the upper limit; so all methods
make a mistake less than 10\% of the time.  Figure \ref{mistakefig} shows the
fraction of mistakes with test (b) using $C_0$ (dotted), $p_{\mathrm{Max}}$
(dash-dotted) and
$C_{\mathrm{Max}}$ (dashed).  Although $C_{\mathrm{Max}}$ and $p_{\mathrm{Max}}$ give equally
strong upper limits for test (b), $C_{\mathrm{Max}}$ makes fewer mistakes.
$C_0$ makes the most mistakes of the tested methods.  Not shown is the Poisson method; because its
upper limit is so high, it makes almost no mistakes.

\section{Conclusions}

Judging from the tests shown in Fig.~\ref{comparefig} and
Fig.~\ref{mistakefig}, the best of the methods discussed here is the
optimum interval method, with $C_{\mathrm{Max}}$.
This method is useful for experiments with small
numbers of events when it is not possible to make an accurate model of the
background, and it can also be used when
experimenters want to show an especially reliable upper limit that doesn't
depend on trusting their ability to model the background.  Because the optimum
interval method automatically avoids parts of the data range in which there
are large backgrounds, it is relatively insensitive to placement of the
cuts limiting the experimental range.  Because the
optimum interval method doesn't use binned data, it cannot be biased by
how experimenters choose to bin their data.  Unlike Bayesian upper limits
with a uniform prior, the result of the optimum interval method is unchanged
when a change in variable is made.
The optimum interval method produces a true, though possibly conservative,
classical (frequentist) confidence interval; at least 90\% of the time the
method is used its 90\% confidence level upper limit will be correct, barring
experimental systematic errors.


\begin{acknowledgments}

Thanks are due to Richard Schnee for useful discussions, for
suggesting improvements of this paper, and for being the first to apply its
methods to an experimental analysis.

This work has been supported by a grant from the U.S.~Department of Energy
No. DE-FG03-91ER40618.

\end{acknowledgments}

\appendix
\section{Derivation of the Equation for $C_0$}

In order to derive Eq.~\ref{C0eqn}, let us first
find the probability that the maximum gap size is less than $x$
when there are exactly $n$ events, then get $C_0$ by
averaging $n$ over a Poisson distribution.

We assume $n$ events are distributed in some variable, $y$,
according to a density distribution that integrates to a total of $\mu$
expected events, and define $P(x;n,\mu)$ to be the probability that the
maximum gap size is less than $x$.
As explained in Section II, one may make a change of variables to $z(y)$
such that the density distribution
is uniform over $0<z<\mu$.  $P(x;n,\mu)$ is the probability that the maximum
$z$ coordinate distance between adjacent events is less than $x$ given that
there are exactly $n$ events distributed randomly, independently, and uniformly
between $z=0$ and $z=\mu$.  The function $P$ depends only on $x$, $n$, and
$\mu$, but not on the shape of the original density distribution.

The problem of finding $P(x;n,\mu)$ can be simplified by making a coordinate
change $w(z) = z/\mu$.  The new coordinate runs from 0 to 1 instead of 0 to
$\mu$.  With this coordinate change, any set of $n$ events
with $x$ equal to the maximum gap between adjacent events becomes a
set of $n$ events, still uniformly distributed, but with maximum new
coordinate distance between adjacent events equal to $x/\mu$.  It follows that
$P(x/\mu;n,1) = P(x;n,\mu)$, and we need only solve the problem of finding $P$
for $\mu=1$ to get the solution for any value of $\mu$.  When $\mu$ is
understood to be 1, it will be dropped, and we will write $P(x;n)$ to mean the
same as $P(x;n,1)$.  The problem has been reduced to one in which $n$ points
have been scattered randomly in independent uniform probability distributions
on the interval $(0,1)$.  We want to find the probability that the maximum
empty interval has length less than $x$.  We do this with the help of a
recursion relation that allows one to compute $P(x;n+1)$ from knowledge of
$P(x;n)$.

$P(x;n+1)$ is the integral over $t<x$ of the probability that the lowest event
is between $t$ and $t+dt$ and that the rest of the $n$ events in the remaining
1-t range has no gap greater than $x$.
The probability that the lowest event is between $t$ and $t+dt$
is (number of ways of choosing one particular event of the $n+1$ events) times
(probability that the particular event will be between $t$ and $t+dt$)
times (probability that each of the other $n$ events will be greater
than $t$).  We get a factor in the integrand $(n+1)\times dt \times (1-t)^n$.
The other factor in the integrand is the probability that
there is no gap greater than $x$ for the remaining $n$ events:
$P(x;n,1-t) = P(x/(1-t);n)$.  The recursion relation for $0<x<1$ is
\be P(x;n+1) = \int_0^x dt\,(n+1)(1-t)^n P\left(\frac{x}{1-t};n\right).
\label{Recursion}\ee

It is convenient to distinguish between various pieces of the $x$ range
between 0 and $\mu$, for it will turn out that $P(x;n,\mu)$ takes on different
forms in different pieces of that range.  If $x$ is in the range
$\mu/(m+1) < x < \mu/m$,
we say $P(x;n,\mu)=P_m(x;n,\mu)$, and we say $x$ is in the $m$'th range.
Let us again restrict ourselves to $\mu=1$ and consider Eq.~\ref{Recursion}.
If $x$ is in the $m$'th range and, as in Eq.~\ref{Recursion},
$0<t<x$, then $x/(1-t)$ is in either range $m$ or range $(m-1)$.
The boundary between these two ranges is at $x/(1-t) = 1/m$; so
$t=1-mx$.  For $m>0$ Eq.~\ref{Recursion} becomes
\bea \frac{P_m(x;n+1)}{n+1} =& \int_0^{1-mx} dt\,(1-t)^n
P_m\left(\frac{x}{1-t};n\right) \nonumber \\
+& \int_{1-mx}^x dt\,(1-t)^n P_{m-1}\left(\frac{x}{1-t};n\right).\nonumber \\
\label{Pmeqn} \eea
The appearance of $m-1$ brings up the question of what happens if $m=0$.  Let
us interpret the $m=0$ range to be the one with $1/1 < x < 1/0=\infty$.  Since the
empty space between events is certainly less than the length of the whole
interval, $P_0(x;n)=1$.

For $m\ge 0$ it can be shown that
\be P_m(x;n) = \sum_{k=0}^m (-1)^k\left({^{n+1} _k}\right) (1-kx)^n.
\label{Pmeqn2}\ee
In this equation, we interpret $\left({^n _k}\right)$ as
$$ \left({^n _k}\right) = \frac{n!}{k! (n-k)!} \equiv \frac{\Gamma(n+1)}
{\Gamma(k+1)\Gamma(n-k+1)}.$$
The gamma function is meaningful when analytically continued, in which
case $\left({^n _k}\right)$ is zero if $k$ is an integer that is less than
zero or greater than $n$.  
In $P(x;0)$, the maximum (and only) gap is always 1; so $P_0(x;0)=1$ for
$x>1$, while for $m>0$, when $0<x<1$, $P_m(x;0) = 0$.
Since Eq.~\ref{Pmeqn2} is easily verified to be
correct for all $m\ge 0$ when $n=0$, one may use induction with
Eq.~\ref{Pmeqn} to
prove Eq.~\ref{Pmeqn2} for all other $n>0$. 
The simple but somewhat tedious manipulations of sums will
not be given here, except for a useful identity in the induction step:
$$\left({^n _k}\right) + \left({^n _{k-1}}\right) = \left({^{n+1} _k}
\right). $$
It follows from Eq.~\ref{Pmeqn2} that
\be P_m(x;n,\mu)= \sum_{k=0}^m (-1)^k\left({^{n+1} _k}\right)
(1-kx/\mu)^n. \label{Pmeqn3} \ee

Let us now compute $C_0$, the probability for the maximum empty space between
events in $(0,\mu)$ being less than $x$ given only that events are thrown
according to a uniform unit density.  Average Eq.~\ref{Pmeqn3} over a
Poisson distribution with mean $\mu$ to get
\be C_0= \sum_{k=0}^m \sum_{n=0}^\infty e^{-\mu} \frac{\mu^n}{n!}(-1)^k
\left({^{n+1} _k}\right)(1-kx/\mu)^n,
\label{C0unsummed} \ee
which can be summed over n (again the manipulations will not be
shown here) to give Eq.~\ref{C0eqn}.

\section{Peculiarities of $\bar C_{\mathrm{Max}}$}

The function $\bar C_{\mathrm{Max}}(.9,\mu)$ has certain peculiarities.  For
example, it cannot be defined for $\mu < 2.3026$.  Random experiments with
$\mu<2.3026$ either give the largest possible value of $C_{\mathrm{Max}}$,
which occurs for zero events, with probability $e^{-\mu}>10\%$, or give smaller
values with probability $1-e^{-\mu}<90\%$.  There is therefore no number,
$\bar C_{\mathrm{Max}}(.9,\mu)$, for which there is exactly 90\% probability of
$C_{\mathrm{Max}}<\bar C_{\mathrm{Max}}(.9,\mu)$.  No cross section resulting
in $\mu<2.3026$ can be excluded to as high a confidence level as 90\%.

Another peculiarity of $\bar C_{\mathrm{Max}}(.9,\mu)$ is that it is not
especially smooth; it tends to increase rapidly near certain
values of $\mu$.  To understand this behavior, note that for a given value of
$\mu$, the maximum possible value of $x$
is $x=\mu$.  Thus the maximum possible value over all $x$ of
$C_n(x,\mu)$ is $C_n(\mu,\mu)$.  If $C_n(\mu,\mu)$ is less than
$\bar C_{\mathrm{Max}}(.9,\mu)$ then intervals with $n$ events cannot have
$C_{\mathrm{Max}}=C_n$ for that value of $\mu$.  Furthermore, since
$C_n(x,\mu)$ decreases with increasing $n$, intervals with $m>n$ events also
have $C_m<C_{\mathrm{Max}}$.  For low enough $\mu$, only intervals
with $n=0$ need be considered.  In this case, the 90\% confidence upper limit
for $C_{\mathrm{Max}}$ occurs when $x$ in $C_0(x,\mu)$ is equal to $x_0(.9,\mu)$,
where $x_0(C,\mu)$ is the inverse of $C_0(x,\mu)$; it is defined as the value
of $x_0$ for which $C_0(x_0,\mu)=C$.  Thus for low enough $\mu$
(but above 2.3026) 
\be \bar C_{\mathrm{Max}}(.9,\mu) = C_0(x_0(.9,\mu),\mu). \label{QMax} \ee
$C_0(x_0(.9,\mu),\mu)=0.9$ from the definitions of $C_0$ and $x_0$.
This formula for $\bar C_{\mathrm{Max}}$ breaks down as soon as $\mu$ is large
enough to have $C_1(\mu,\mu) > \bar C_{\mathrm{Max}}(.9,\mu)$, for
at this value of $\mu$ it is possible for an interval with $n=1$ to be
$C_{\mathrm{Max}}$.  In general, the threshold $\mu$ for intervals with $n$ points
being able to produce $C_{\mathrm{Max}}$ for confidence level $C$ is where
\be C_n(\mu,\mu) = \bar C_{\mathrm{Max}}(C,\mu). \label{QNmumu} \ee
Every time a threshold in $\mu$ is passed that allows another value of $n$
to participate in producing $C_{\mathrm{Max}}$, the value
of $\bar C_{\mathrm{Max}}(C,\mu)$ spurts upward.

If one considers all intervals with $\le n$ events, then the largest expected
number of events is
less than $\mu$ if and only if there are more than $n$ events in the entire
experimental range.  Thus $C_n(\mu,\mu)$ is the probability of $>n$ events
in the entire experimental range: $C_n(\mu,\mu)=P(\mu,n+1)$ of
Eq.~\ref{PoisProb}.
This equation, with Eq.~\ref{QNmumu}, can be used to
compute the thresholds
in $\mu$ where $n$ events first need to be included when trying to find
$C_{\mathrm{Max}}$ in a calculation of the 90\% confidence level.  These
thresholds are tabulated in table~\ref{muCvsNtab}.  As an example of usage
of this table, if you are evaluating $C_{\mathrm{Max}}$ for a 90\%
confidence level calculation with $\mu=20$, you can ignore intervals with
more than 11 events.

\begin{table}
\caption{Threshold $\mu$ for which intervals with $\ge n$ events need
not be considered when computing $C_{\mathrm{Max}}$.}
\label{muCvsNtab}
\begin{tabular}{|r||r|r|r|r|r|}
\hline
n  & $\mu$(n) & $\mu$(n+1) & $\mu$(n+2) & $\mu$(n+3) & $\mu$(n+4) \\
\hline
    0  &   2.303  &   3.890  &   5.800  &   7.491  &   9.059 \\
    5  &  10.548  &  12.009  &  13.433  &  14.824  &  16.196 \\
   10  &  17.540  &  18.891  &  20.208  &  21.520  &  22.821 \\
   15  &  24.119  &  25.400  &  26.669  &  27.926  &  29.197 \\
   20  &  30.457  &  31.690  &  32.972  &  34.203  &  35.422 \\
   25  &  36.632  &  37.849  &  39.108  &  40.333  &  41.546 \\
   30  &  42.768  &  43.978  &  45.164  &  46.351  &  47.544 \\
   35  &  48.734  &  49.944  &  51.139  &  52.314  &  53.488 \\
\hline
\end{tabular}
\end{table}

The many rapid
increases in $\bar C_{\mathrm{Max}}(.9,\mu)$ of Fig.~\ref{CMaxfig} occur
when thresholds given in table~\ref{muCvsNtab} are crossed.

\section{Probability of More Events Than Observed in an Interval}

Instead of using $C_n(x,\mu)$ as a measure of how strongly a given interval
with $n$ events excludes a given cross section, one may use
$p_n(x)$, the calculated Poisson probability of there being more events
in a random interval of that size than were actually observed.  This
probability is $P(x,n+1)$, as defined in Eq.~\ref{PoisProb}.
If $p_n$ is too large, then the
cross section used in the calculation must have been too large.
For a given cross section, find the
interval that excludes the cross section most strongly; i.e., find the
interval that gives the largest calculated probability of there being more
events in the interval than were actually observed.  In other words,
as was done with $C_{\mathrm{Max}}$ of the optimum interval method, define
$p_{\mathrm{Max}}$ to be the maximum over the $p_n$ for all possible
intervals.  If random experiments for the same given cross section would
give a smaller $p_{\mathrm{Max}}$ 90\% of the
time, then the cross section is rejected as too high with 90\% confidence
level.  The function, $\bar p_{\mathrm{Max}}(C,\mu)$, is defined as the
$p_{\mathrm{Max}}$ for which confidence level $C$ is reached at the given
$\mu$.

Although this method
may not be as effective as the optimum gap method, it is much
easier to calculate $p_n(x)=P(x,n+1)$ than it is to calculate $C_n(x,\mu)$.

Much of the reasoning applied to the optimum interval method applies here.
As was the case for the optimum interval method,
$\bar p_{\mathrm{Max}}(C,\mu)$ depends only on $C$ and $\mu$, but not
otherwise on the shape of the cross section.  As for the optimum interval
method, $\bar p_{\mathrm{Max}}(.9,\mu)$ is not defined for $\mu<2.3026$.
For sufficiently low $\mu$ above 2.3026 Eq.~\ref{QMax} becomes
\be \bar p_{\mathrm{Max}}(.9,\mu) = p_0(x_0(.9,\mu)) = e^{-x_0(.9,\mu)}. \label{QMaxp} \ee
For the threshold $\mu$ at which intervals with $n$ points become
able to contribute to $p_{\mathrm{Max}}$ for confidence level $C$,
Eq.~\ref{QNmumu} becomes
\be P(\mu,n+1) \equiv p_n(\mu) = \bar p_{\mathrm{Max}}(C,\mu). \label{QNmumup} \ee

\begin{figure}
\includegraphics{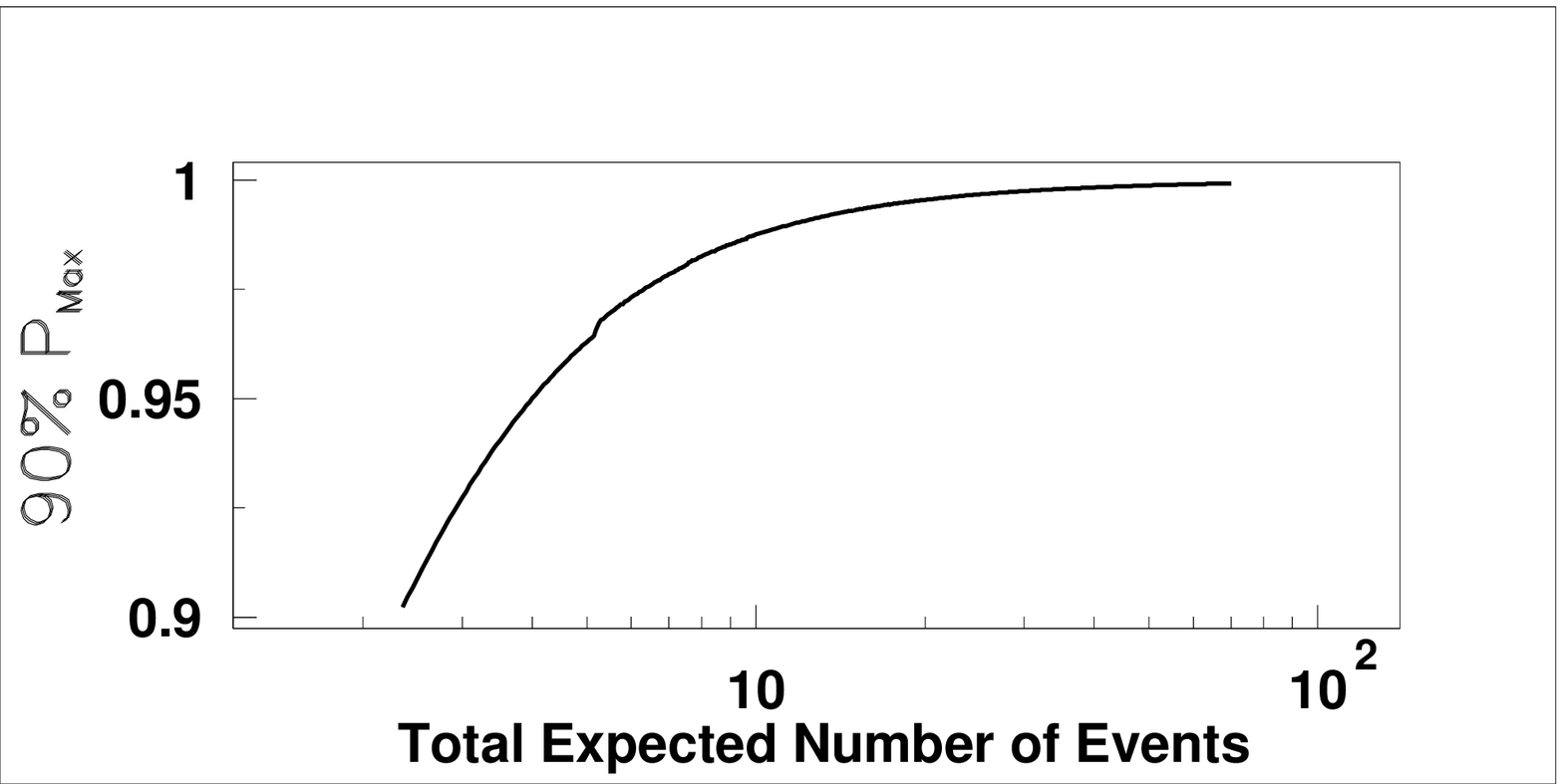}
\vskip 1mm
\caption{Plot of $\bar p_{\mathrm{Max}}(.9,\mu)$, the value of
$p_{\mathrm{Max}}$ for which the 90\% confidence level is reached, as a
function of the total number of events $\mu$ expected in the experimental
range.}
\label{BMaxfig}
\end{figure}

A Monte Carlo program was used to compute a table of
$\bar p_{\mathrm{Max}}(0.9,\mu)$ for $\mu\le 70$, and the function is plotted
in Fig.~\ref{BMaxfig}.

\begin{table}
\caption{Threshold $\mu$ below which intervals with $\ge n$ events
need not be considered when computing $p_{max}$ for the 90\% confidence
level.}
\label{muBvsNtab}
\begin{tabular}{|r||r|r|r|r|r|}
\hline
n  & $\mu$(n) & $\mu$(n+1) & $\mu$(n+2) & $\mu$(n+3) & $\mu$(n+4) \\
\hline
    0  &   2.303  &   5.156  &   7.584  &   9.661  &  11.599 \\
    5  &  13.427  &  15.193  &  16.900  &  18.559  &  20.176 \\
   10  &  21.771  &  23.355  &  24.880  &  26.419  &  27.922 \\
   15  &  29.428  &  30.891  &  32.359  &  33.808  &  35.251 \\
   20  &  36.701  &  38.100  &  39.519  &  40.913  &  42.317 \\
   25  &  43.700  &  45.091  &  46.465  &  47.827  &  49.193 \\
   30  &  50.561  &  51.902  &  53.255  &  54.589  &  55.926 \\
   35  &  57.264  &  58.603  &  59.920  &  61.237  &  62.549 \\
   40  &  63.868  &  65.179  &  66.478  &  67.791  &  69.080 \\
\hline
\end{tabular}
\end{table}

Table~\ref{muBvsNtab} shows approximate
values of the threshold $\mu$ calculated using Eq.~\ref{QNmumup} with $C=0.9$
for each $n$ from 0 to 44.  The third digit of
$\mu$ does not really deserve to be trusted since $\bar p_{\mathrm{Max}}$ was
computed from a Monte Carlo generated table.

Appendix B explained why the value of $\bar C_{\mathrm{Max}}(.9,\mu)$ spurts
upward when $\mu$ crosses a threshold where intervals with more points can
contribute to $C_{\mathrm{Max}}$.  A much less obvious similar effect
occurs with $\bar p_{\mathrm{Max}}(.9,\mu)$.  Notice the
irregularity in the curve of Fig.~\ref{BMaxfig} just after $\mu=5.156$, where
$n=1$ first begins to contribute.  Between $\mu=2.3026$ and $\mu=5.156$,
Eq.~\ref{QMaxp} applies,
but after $\mu=5.156$, $\bar p_{\mathrm{Max}}$ shoots above this form.
The smaller irregularity above $\mu=7.584$, where $n=2$ begins to contribute,
is barely visible.

\bibliographystyle{prsty}

\begin{thebibliography}{10}

\bibitem{FeldCous} G.~J. {Feldman} and R.~D. {Cousins}, Phys. Rev. D
{\bf 57},  3873  (1998).

\bibitem{CDMS} D. Abrams, {\it et~al.} (2002), submitted to
Phys. Rev. D, astro-ph/0203500.


\bibitem{software} http://www.slac.stanford.edu/$\sim$yellin/ULsoftware.html

\bibitem{PDG}
Particle Data Group, Phys. Rev. {\bf D54} 1 (1996).  See especially page 164.
In subsequent reviews the Particle Data Group dropped its discussion of upper
limits of Poisson processes,
probably because the procedure of Feldman and Cousins~\cite{FeldCous}
is now generally accepted as preferable when, as is usually assumed in
discussions of confidence regions, backgrounds are well understood.

\bibitem{CERNLIB} http://wwwinfo.cern.ch/asdoc/shortwrupsdir/index.html

\end{thebibliography}

\end{document}